\documentclass[prl,twocolumn,superscriptaddress,showpacs,amsmath,amssymb,amsfonts]{revtex4-1}
\usepackage[dvipdfmx]{graphicx,color}
\usepackage{subfigure}
\usepackage{bm}
\usepackage{siunitx}
\usepackage{amsthm,mathtools}

\def\U#1{{\rm #1}} 
\def\u#1{_{\rm #1}}
\newcommand{\ket}[1]{| #1 \rangle}

\newcommand{\expect}[1]{\left\langle #1 \right\rangle} 
  
\newcommand{\vect}[1]{\boldsymbol{#1}}

\newcommand{\gn}[1]{g^{(#1)}}
\newcommand{\Vn}[1]{V^{(#1)}}

\def\PQ{P\u{coinc}^{\U{id}}}
\def\PC{P\u{coinc}^{\U{dist}}}
\def\PQT{\tilde{P}\u{coinc}^{\U{id}}}
\def\PCT{\tilde{P}\u{coinc}^{\U{dist}}}

\begin{document}
\title{
  Statistical-noise-enhanced multi-photon interference
}
\author{Rikizo Ikuta}
\email{ikuta.rikizo.es@osaka-u.ac.jp}
\affiliation{
  \mbox{Graduate School of Engineering Science, The University of Osaka,
  Osaka 560-8531, Japan}}
\affiliation{
  \mbox{Center for Quantum Information and Quantum Biology, The University of Osaka,
  Osaka 560-0043, Japan}}

\begin{abstract}
  Photon statistics plays a governing role in multi-photon interference.
  While interference visibility in the standard two-photon case, known as Hong-Ou-Mandel interference,
  monotonically degrades with higher intensity correlation functions,
  we show that this monotonicity does not hold for three-photon interference in symmetric circuits. 
  We reveal that, in the discrete Fourier transform circuit,
  engineered super-Poissonian photon-number fluctuations,
  realized using a modulated laser, maximize the visibility,
  surpassing the magnitude of the single-photon signature.
  In addition, by tuning the symmetric circuit parameters,
  we demonstrate that the visibility hierarchy inverts relative to the benchmark of Poissonian statistics. 
  This trade-off implies that quantum and classical advantages are mutually exclusive resources 
  for interference, indicating a form of statistical complementarity.
\end{abstract}

\maketitle

{\it Introduction. -- }
Two-photon interference, represented by the Hong-Ou-Mandel~(HOM) effect~\cite{Hong1987}, is fundamentally characterized by boson bunching, observed as the suppression of coincidence events. This phenomenon extends to multi-photon linear optical circuits, where interference is dictated by permanents of a transformation matrix, leading to intricate behaviors such as generalized bunching and zero-transmission laws~\cite{Lim2005,Tichy2010,Spagnolo2013-2,Crespi2015,Shchesnovich2016,Crespi2016,Dittel2018,Yamazaki2023arxiv}.
These developments have naturally fostered the intuition that perfectly indistinguishable photons represent the ultimate resource for maximizing interference contrast.
This perspective is primarily shaped by studies of partial distinguishability or mode mismatch~\cite{Shchesnovich2015,Tichy2015,Karczewski2019,Shchesnovich2022,Annoni2025}. 
Recent studies in this domain have revealed that bunching probabilities can exhibit highly non-trivial
and non-monotonic dependencies on the distinguishability parameter~\cite{Seron2023,Pioge2023},
that governs the transition from quantum indistinguishability to classical distinguishability, 
reflecting the rich structure of multi-photon interference.

Crucially, however, mode mismatch is not the sole origin of classicality. 
Light possesses another fundamental degree of freedom, namely photon statistics~\cite{Glauber1963,Mandel1995,Klyshko1996}.
This degree of freedom is characterized by the normalized autocorrelation functions 
$\gn{n}=\expect{:\hat{n}^n:}/\expect{\hat{n}}^n$,
defined by the normally ordered moments of the photon-number operator $\hat{n}$. 
It parametrizes a distinct axis of the quantum-classical transition, typically characterized by $\gn{2}$.
Complementing the extensively studied distinguishability axis,
this approach allows us to explore the expansive state space of unbounded photon numbers,
unveiling interference phenomena that remain inaccessible in analyses restricted to fixed-photon-number states.

Investigating this second axis requires a distinct methodological approach.
While the effects of partial distinguishability are often analyzed through raw coincidence probabilities,
comparing these raw rates across different statistical sources is potentially misleading.
Unlike single-photon states, general light sources spanning the sub-Poissonian to super-Poissonian regimes 
inherently contain multi-photon components that trivially scale the background coincidence rates.
In particular, super-Poissonian light with $\gn{2} > 1$ exhibits enhanced coincidence probabilities
due to classical bunching, independent of any interference effect.
Therefore, to isolate genuine interference signatures from this trivial statistical scaling,
we focus on the interference visibility,
defined as the normalized contrast between the fully indistinguishable and distinguishable coincidence rates.

In this paper,
we demonstrate that statistical noise associated with photon-number fluctuations does not necessarily degrade interference contrast. 
While the HOM effect implies a monotonic reduction of visibility with increasing $\gn{2}$~\cite{Santori2002,Tsujimoto2017,Ollivier2021,Tsujimoto2023},
we show that this intuition does not hold for general multiport interference.
Focusing on symmetric unitary transformations, including the discrete Fourier transform~(DFT),
we reveal that appropriate statistical noise can enhance the interference contrast 
beyond the single-photon limit. Furthermore, we show that the visibility hierarchy 
among sub-Poissonian, Poissonian, and super-Poissonian lights is not universal 
but depends on the specific unitary configuration.
By employing Fock-state and engineered-noise inputs, 
we find that the resources associated with sub-Poissonian and super-Poissonian statistics 
cannot be effectively utilized simultaneously for interference,
suggesting an underlying statistical complementarity.
This qualitative inversion of the quantum-classical hierarchy clarifies that 
the statistical nature of the input field acts not merely as background noise,
but as a tunable control parameter that can enhance interference capabilities.

\begin{figure}[t]
 \begin{center}
   \scalebox{0.7}{\includegraphics{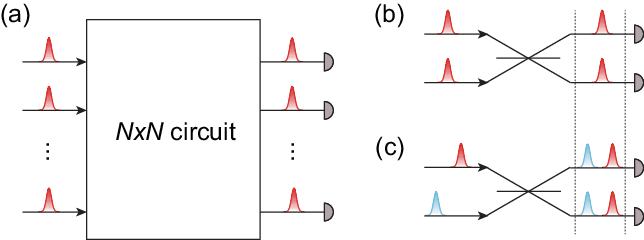}}
   \caption{
     (a) Schematic of the interferometric scenario.
     The input fields are not limited to single photons
     but are characterized by intensity correlation functions. 
     (b, c) Conceptual illustration of the coincidence probabilities in HOM interference 
     (b) with the inputs of indistinguishable photons~($\PQ$) and  
     (c) those of distinguishable photons ($\PC$). 
   }
    \label{fig:circuit}
 \end{center}
\end{figure}
{\it Framework for multi-photon interference. -- }
We first introduce a framework for analyzing multi-photon interference as shown in Fig.~\ref{fig:circuit}~(a).  
We evaluate an interference visibility based on the $N$-fold coincidence detection probability $P\u{coinc}$, 
defined by the normally ordered photon-number correlation up to a proportionality constant, as 
\begin{align}
  P\u{coinc} = \expect{:\hat{n}^{\U{out}}_1 \cdots \hat{n}^{\U{out}}_N:}, 
\end{align}
where $\hat{n}^{\U{out}}_i$ is the photon number operator for output mode $i$. 
We denote the coincidence probabilities for perfectly indistinguishable
and completely distinguishable photons entering different input ports by $\PQ$ and $\PC$, respectively.
While detection probabilities are not strictly proportional to photon numbers, 
particularly for unit-efficiency detectors, 
this description is valid in the low detection-probability regime relevant to linear optical experiments~\cite{Santori2002,Tsujimoto2017,Ollivier2021,Tsujimoto2023}. 
Moreover, the photon-number formalism offers a detector-independent perspective 
for analyzing the influence of photon statistics 
on multi-photon interference effects~\cite{Glauber1963,Ou1987,Ikuta2022,Tsujimoto2023}.

The visibility in this paper is an extension of the one used in HOM experiments.
It is defined by normalizing the interference contrast
against the distinguishable limit of the same light source, given by
\begin{align}
    \Vn{n} = 1 - \frac{\PQ}{\PC}. 
    \label{eq:vis}
\end{align}
As a natural extension of HOM interference~\cite{Hong1987} to multi-photon interference, 
we assume that the input fields are statistically independent and have no phase correlations. 
For simplicity, 
we assume that the light field entering each input port occupies a single mode, 
described by a single creation operator, regardless of its photon statistics. 
Furthermore, we assume symmetric inputs with identical mean photon number
and photon statistics $\gn{N},\ldots,\gn{2}$ for all input modes.
This simple configuration is sufficient
to reveal the rich structure of multi-photon interference considered here.
For notational brevity, we define $\tilde{P}\u{coinc}$ as the coincidence probability
normalized by the input intensities, as the absolute photon number scaling does not affect the visibility
for the symmetric inputs. 

The visibility defined in Eq.~(\ref{eq:vis})
quantifies the intrinsic interference capability of the light source. 
Since the definition effectively cancels the dominant statistical scaling
associated with the non-interfering background contributions, 
the interference contrast is isolated from trivial intensity fluctuations.
Unless otherwise stated, we use this definition throughout our main results.
We note that, as an alternative, one may normalize the coincidence probability 
by a reference probability obtained from uncorrelated components~\cite{Rarity2005} as 
$1 - \PQ/(\expect{\hat{n}^{\U{out}}_1}\cdots\expect{\hat{n}^{\U{out}}_n})$. 
However, such a definition introduces a trivial background scaling factor $g^{(N)} > 1$, 
which renders the resulting visibility largely insensitive to genuine interference effects
and thereby degrades its usefulness as a meaningful measure of interference.

{\it Two-Photon Interference. -- }
Before analyzing multi-photon circuits, 
we revisit the HOM interference as shown in Figs.~\ref{fig:circuit}~(b) and (c).
We consider a $2 \times 2$ beamsplitter with reflectance $R$ and transmittance $T$ satisfying $R+T=1$.
The coincidence probabilities for indistinguishable and distinguishable inputs are given by 
\begin{align}
  \PQT &= 1 - 2 RT (2-\gn{2}),
        \label{eq:PQ2}
  \\
  \PCT &= 1 - 2 RT (1-\gn{2}).
        \label{eq:PC2}
\end{align}
Here, the gap $2RT$ represents the interferometric effect
between indistinguishable single photons incident from different input ports.
Substituting these into Eq.~(\ref{eq:vis}), we obtain the explicit form of the visibility as 
\begin{equation}
  \Vn{2} = \frac{2RT}{2RT \gn{2} + 1-2RT}. 
    \label{eq:V2}
\end{equation}
Since $0 \leq 2RT \leq 0.5$, ensuring that $\Vn{2}\geq 0$, we obtain
$\partial\Vn{2}/\partial{\gn{2}}=-(\Vn{2})^2 \leq 0$. 
This means that $\Vn{2}$ is a strictly monotonically decreasing function of statistical noise $\gn{2}$,
indicating that increasing the noise unilaterally degrades visibility
regardless of the quantum or classical regime.
This supports the intuition that quantum states are the better resource,
as the single-photon inputs yields the maximum visibility.

{\it Noise-Enhanced visibility in 3-port beamsplitter. -- }
We demonstrate that, in contrast to the $N=2$ case,
single-photon inputs do not necessarily yield the largest interference visibility magnitude for $N=3$,
as exemplified by the DFT circuit.
The transformation matrix is described by 
\begin{align}
  U\u{DFT}=\frac{1}{\sqrt{3}}
  \begin{pmatrix}
    1 & 1 & 1\\
    1 & \omega & \omega^2\\
    1 & \omega^2 & \omega
  \end{pmatrix},
\label{eq:DFT}
\end{align}
where $\omega=e^{2i\pi/3}$. 
The coincidence probabilities are given by using $\gn{3}$ and $\gn{2}$ as~\cite{SM} 
\begin{align}
  \PQT &= \frac{1}{9} g^{(3)} + \frac{1}{3},
        \label{eq:PQDFT}\\ 
  \PCT &= \frac{1}{9} g^{(3)} + \frac{2}{3} g^{(2)} + \frac{2}{9}. 
       \label{eq:PCDFT}
\end{align}
In both equations, the last terms represent the coincidence probabilities
for single-photon inputs.
In $\PQT$, the $g^{(2)}$ term  vanishes as a result of interference. 
Substituting these into $\Vn{3}$ in Eq.~(\ref{eq:vis}), we obtain
\begin{equation}
  \Vn{3} = \frac{6g^{(2)} - 1}{g^{(3)} + 6g^{(2)} + 2}. 
\label{eq:V3}
\end{equation}

Different from the case of $N=2$, 
$\Vn{3}$ takes negative values in the nonclassical region of $\gn{2} < 1/6$, 
reaching a minimal value of $-0.5$ for the single-photon inputs. 
A negative visibility corresponds to a peak~(bump) 
in the coincidence probability for the inputs of indistinguishable photons,
relative to distinguishable photons. 
This behavior originates from the constructive interferometric effect of single photons, 
each satisfying $\gn{3}=\gn{2}=0$, rather than from higher-photon-number components. 

\begin{figure}[t]
 \begin{center}
   \scalebox{1.3}{\includegraphics{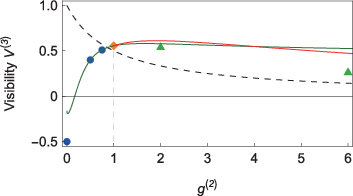}}
   \caption{
     Visibility $\Vn{3}$ in DFT circuit.
     The red solid curve indicates the upper bound $\Vn{3}\u{cl}$
     in the classical regime with $\gn{2}\geq 1$.
     The green solid curve represents the limit of $\Vn{3}$ for the pure Gaussian-state inputs. 
     For reference, $\Vn{2}=(1+\gn{2})^{-1}$ with $T=R=1/2$ in Eq.~(\ref{eq:V2}) 
     is shown as the black dashed curve.
     The plotted points represent, in ascending order of $\gn{2}$, 
     inputs with 1, 2, and 4 photons~($\circ$)~\cite{SM}, laser light~($\diamond$), thermal ($\gn{2}=2, \gn{3}=6$)
     and its second-harmonic ($\gn{2}=6, \gn{3}=90$) lights~({\scalebox{0.7}{$\triangle$}})~\cite{Loudon2000}
     in each mode. 
   }
   \label{fig:DFT_vis}
 \end{center}
\end{figure}
While a negative visibility is a noteworthy feature, 
it can be inferred from previous reports~\cite{Tichy2010, Spagnolo2013-2, Menssen2017}. 
The focal point of this study, however, is the counter-intuitive behavior observed in the classical regime.
From the Cauchy–Schwarz inequality $\expect{X^2}\expect{Y^2} \geq \expect{XY}^2$ 
with $X=n^{3/2}$ and $Y=n^{1/2}$, $\gn{3}\geq (\gn{2})^2$ is satisfied in the classical wave theory. 
This leads to the upper bound on $\Vn{3}$ as
\begin{align}
  \Vn{3}\u{cl} = \frac{6g^{(2)} - 1}{(\gn{2})^2 + 6\gn{2} + 2} \geq \Vn{3}. 
  \label{eq:Vcl}
\end{align}
The condition $\gn{3} = (\gn{2})^2$ for achieving the upper bound is satisfied 
by a classical mixture of laser light satisfying $\gn{2}=\gn{3}=1$
at a fixed non-zero intensity with probability $p$ 
and the vacuum with probability $1-p$.
This mixture gives $\gn{3}=p^{-2}$ and $\gn{2}=p^{-1}$.

In Fig.~\ref{fig:DFT_vis}, we plot the dependence of $\Vn{3}\u{cl}$ on $\gn{2}$, 
along with the visibilities $\Vn{3}$ for the representive input configurations. 
We see that, different from the monotonic decay $\Vn{2}$ for $N=2$, 
$\Vn{3}\u{cl}$ is non-monotonic with respect to $g^{(2)}$
and retains relatively large value even as $g^{(2)}$ increases. 
From Eq.~(\ref{eq:Vcl}), 
the maximum of $\Vn{3}\u{cl}$ is $(19-\sqrt{109})/14 \sim 0.61$,
achieved by the above mixed state with $p=6/(1+\sqrt{109})\sim 0.52$, $\gn{2}=p^{-1} \sim 1.9$ and $\gn{3} \sim 3.6$.
$\Vn{3}$ decreases beyond its maximum, tending to zero as $\gn{2}\rightarrow \infty$.
We note that for a laser light and for thermal light~($\gn{2}=2$, $\gn{3}=6$),
the corresponding $\Vn{3}$ values are $5/9\sim 0.56$ and $11/20\sim 0.55$, respectively,
which are close to the maximum.

The results reveal counter-intuitive phenomena that challenge the conventional hierarchy of quantum optics. 
First, we observe a qualitative inversion of the interference signature. 
While the pure single photons result in a probability peak and negative visibility, 
classical inputs form a dip~($\Vn{3} > 0$) due to destructive interference. 
The maximum value $\Vn{3} \sim 0.61$ in the classical regime exceeds the magnitude of the visibility $|\Vn{3}| = 0.5$ obtained for single-photon inputs with $\gn{2}=0$ in each input mode,
thereby establishing a clear separation that goes beyond the conventional expectation of quantum superiority.
Second, this superior contrast is achieved by explicitly introducing statistical noise. 
The maximum visibility is attained not by a stable laser source,
but rather by a classical mixture exhibiting partial bunching with
$\gn{2}\sim 1.9$ and $\gn{3}\sim 3.6$, as simulated using a modulated laser. 
This demonstrates that the statistical noise can act as a vital resource to sharpen the interference,
when carefully tuned to balance the trade-off against the higher-order $\gn{3}$ penalty. 

In addition to its fundamental interest,
our results offer a highly practical advantage for circuit characterization. 
In HOM experiments, classical light sources are often used for alignment, 
in which a higher coincidence flux with $\gn{2}>1$ significantly reduces $\Vn{2}$
as shown in Fig.~\ref{fig:DFT_vis}. 
However, in the DFT setup for $N=3$, 
using an appropriate super-Poissonian light allows one to avoid this trade-off constraint. It provides higher count rates and a sharper interference signature, making it particularly useful for calibration tasks, such as adjusting the timing of input photons, without consuming valuable and fragile quantum resources.

We remark that 
the above analysis focused on the visibilities for classical light and single-photon inputs.
If arbitrary quantum states are allowed, such as mixture of the vacuum, $\ket{1}$, and $\ket{2}$ states, 
visibilities spanning the full range from $-0.5$ to $1$ can be realized~\cite{SM}.
However, such quantum states are not practical resources in standard quantum optics experiments.
Moreover, it is precisely the statistical fluctuations~(noise) that enhance the visibility in these cases.
This stands in stark contrast to standard interferometers, where noise is typically understood to degrade interference contrast.

It is also instructive to compare our results with the recent theoretical bounds 
derived for Gaussian states~\cite{Hotter2025}, 
where it has been shown that pure Gaussian states must satisfy an inequality as 
\begin{align}
  \gn{3} \geq (2-3\sqrt{\gn{2}})^2. 
  \label{eq:gauss}
\end{align}
The corresponding upper~(lower) bound on the visibility for $\Vn{3}\geq 0$~($\Vn{3} < 0$)
derived from this inequality is plotted in Fig.~\ref{fig:DFT_vis}.
This comparison reveals several interesting features. 
First, the maximum visibility for pure Gaussian states is $\sim 0.58$ at $\gn{2} \sim 1.7$,
which is strictly smaller than the value $\sim 0.61$ achievable with classical light. 
Second, and remarkably, from Ref.~\cite{Hotter2025}, 
mixtures of Gaussian states in the regime $\gn{2} > 4/9$ can violate the bound in Eq.~(\ref{eq:gauss}). 
This implies that appropriately engineered photon-number noise can allow for a visibility higher than that of pure Gaussian states. 
Third, the upper bound converges to a finite value of $0.4$ in the limit $\gn{2}\rightarrow \infty$, 
indicating that a robust visibility persists even for arbitrarily large values of $\gn{2}$. 
Fourth, in the regime of $\gn{2}\leq 4/9$, any measured $\Vn{3}$ 
lying outside this bound serves as a witness of non-Gaussianity.

\begin{figure}[t]
 \begin{center}
   \scalebox{1.3}{\includegraphics{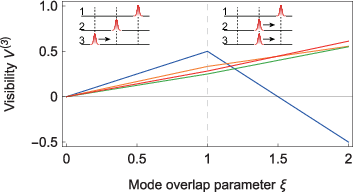}}
    \caption{
      Visibility as a function of the sequential mode overlap parameter $\xi$. 
      The range $0 \leq \xi \leq 1$ corresponds to sweeping the mode overlap $M_{23}$ from 0 to 1
      while keeping $M_{12}=M_{31}=0$.
      The range $1 < \xi \leq 2$ corresponds to sweeping $M_{12}=M_{31}$ from 0 to 1 with $M_{23}=1$.
      Blue: single photon. Orange: laser light. Green: thermal light.  
      Red: optimized noise~($\gn{2}\sim 1.9$, $\gn{3}\sim 3.6$). 
    }
    \label{fig:modemismatch}
 \end{center}
\end{figure}
So far, we have used the probability $\PQT$ calculated under the assumption of perfect mode matching, 
to evaluate the visibility. 
In the final part of this analysis, 
we examine the dependence of the visibility on mode matching considering photon statistics. 
Unlike the case of $N=2$, the notion of mode matching is not uniquely defined, 
since the pairwise mode overlaps $M_{ij}$ between the input modes $i$ and $j$ are not necessarily equal. 
Motivated by an experimental feasibility, we assume a sequential alignment of 
the three mode overlaps for a natural and intuitive framework, like in Fig.~\ref{fig:modemismatch}. 
Under this setting, we derive the corresponding probabilities as~\cite{SM} 
\begin{align}
  \tilde{P}\u{coinc}^{(0,M,0)} &= \frac{1}{9}\gn{3} + \frac{2(3-M)}{9} \gn{2} + \frac{2-M}{9},
                         \label{eq:P1}\\
  \tilde{P}\u{coinc}^{(M,1,M)} &= \frac{1}{9}\gn{3} + \frac{4(1-M)}{9} \gn{2} + \frac{1+2M}{9},  
                         \label{eq:P2}
\end{align}
where the superscripts show values of $(M_{12},M_{23},M_{31})$.
The equations satisfy $\tilde{P}\u{coinc}^{(0,0,0)}=\PCT$ and $\tilde{P}\u{coinc}^{(1,1,1)}=\PQT$. 
Figure~\ref{fig:modemismatch} illustrates the evolution of the visibilities 
calculated using Eqs.~(\ref{eq:P1}) and (\ref{eq:P2}) instead of $\PQT$
as a function of the sequential mode overlap parameter $\xi$. 
In the interval $0\leq \xi\leq 1$, where the photons in mode 1 remain distinguishable,
the system effectively operates in the pairwise two-photon interference regime. 
The visibility behaves analogously to the HOM dip, 
showing a linear dependence on the overlap~\cite{Tsujimoto2023}
and monotonic dependence of $\gn{2}$. 
However, a distinct transition occurs in the range $1\leq \xi\leq 2$.
As the third photon becomes indistinguishable,
the physics shifts from pairwise interactions to genuine three-photon interference. 
Crucially, as the system approaches perfect mode matching ($\xi \to 2$),
the visibility curves exhibit a crossover,
culminating in the non-trivial ordering of visibilities, where noise enhances the signal, 
as predicted in Eq.~(\ref{eq:V3}) and Fig.~\ref{fig:DFT_vis}.

{\it Circuit-dependent reordering of visibilities. -- }
While the DFT provides a paradigmatic example of non-monotonicity,
it represents only a single point in the space of unitary transformations.
To explore a richer interference landscape inherent in three-photon dynamics,
we introduce a symmetric unitary matrix defined by 
\begin{align}
  U\u{sym} = 
  \begin{pmatrix}
    \alpha & \beta  & \beta\\
    \beta  & \alpha & \beta\\
    \beta  & \beta  & \alpha
  \end{pmatrix},
  \label{eq:sym}
\end{align}
where $\alpha = (2+e^{i\phi})/3$ and $\beta = (-1+e^{i\phi})/3$,
with a controllable phase parameter $\phi$. 
In this parameterization, $\phi = 0$ corresponds to the identity operation,
while $\phi = 2\pi/3$ and $\phi = 4\pi/3$
yield a balanced splitting ratio~($|\alpha|=|\beta|=1/\sqrt{3}$),
equivalent to the DFT matrix.

\begin{figure}[t]
 \begin{center}
   \scalebox{1.3}{\includegraphics{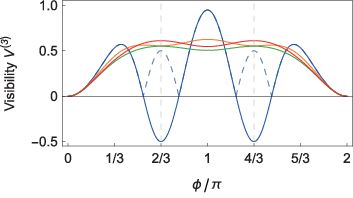}}
    \caption{
      Visibility as a function of phase $\phi$ in $U\u{sym}$.
      Blue: single photon~(the dashed curve indicates the absolute value). 
      Orange: laser light. Green: thermal light. Red: optimized noise for the DFT. 
    }
    \label{fig:sym_vis}
 \end{center}
\end{figure}
Using the coincidence probabilities $\PQT (\phi)$ and $\PCT (\phi)$ 
obtained for fully indistinguishable inputs and for fully distinguishable inputs, respectively, 
we obtain a visibility $\Vn{3}(\phi)=1-\PQT (\phi)/\PCT (\phi)$ as a function of $\phi$~\cite{SM}. 
In Fig.~\ref{fig:sym_vis},
we compare the phase dependence of $\Vn{3}(\phi)$ for different light sources. 
At $\phi=0$ and $2\pi$, the identity $U\u{sym}$ yields zero visibility for all sources. 
At the DFT points, the visibility hierarchy in Fig.~\ref{fig:DFT_vis} is recovered. 
The graph reveals a striking contrast in sensitivity.
The single-photon visibility~(blue) is highly sensitive to the phase, 
oscillating between positive and negative values, 
corresponding to bunching and anti-bunching signatures, respectively, 
and reaching a near-unity magnitude~($\sim 0.95$) at $\phi=\pi$. 
In contrast, the classical light sources show a much smoother variation with the phase.
As a result, the hierarchy of visibilities is not fixed but phase-dependent.
The crossover of the curves demonstrates that
large autocorrelation functions~($g^{(2)}, g^{(3)}$) are not inherently adverse to interference.
Instead, under specific phase conditions such as the DFT,
appropriate noise can be harnessed to enhance the visibility. 

Interestingly, taking the laser-light inputs with $\gn{2} = 1$ as a benchmark, 
we never observe a regime in which this benchmark yields the lowest visibility. 
Notably, this classical advantage is not accessible with standard thermal light~($\gn{2}=2$, $\gn{3}=6$)
but requires engineered intensity fluctuations. 
This implies that the distinct quantum advantage relying on lower $\gn{2}$ 
and the classical noise advantage relying on engineered higher $\gn{2}$ 
constitute mutually exclusive resources for this interference, 
which can be interpreted as statistical complementarity. 
We note that numerical analysis reveals a narrow phase window 
where both the Fock-state~($n\geq 3$) and engineered-noise inputs simultaneously
surpass the laser visibility limit, but only by a negligible amount~($\sim 10^{-3}$)~\cite{SM}. 

{\it Conclusion. -- }
In conclusion, we have investigated the role of photon statistics in multi-photon interference within symmetric unitary circuits. 
In contrast to HOM interference, we find that the multi-photon visibility exhibits a non-monotonic dependence on intensity correlation functions.
Crucially, classical lights and Gaussian states can achieve higher visibility than single-photon inputs in the DFT circuit, with the maximum attained through engineered super-Poissonian fluctuations.
This high visibility accessible with classical light provides a practical advantage for efficient alignment and calibration of multi-photon circuits without fragile quantum resources. 
By tuning the phase parameter of the symmetric circuit in Eq.~(\ref{eq:sym}),
we further showed that the visibility hierarchy inverts relative to the Poissonian baseline. 
This trade-off indicates that quantum and classical advantages are mutually exclusive resources, 
highlighting the concept of statistical complementarity. 

Our work also opens new avenues for exploring the rich landscape of multi-photon interference 
along the axis of photon statistics bridging the quantum and classical domains.
While we focused here on the minimal extension to the three-photon case restricted to symmetric configurations,
the observed statistical advantage is likely circuit-dependent.
For example, in the four-photon case~($N=4$),
the suppression of specific output events governed by zero-transmission laws
may restore the advantage of sub-Poissonian statistics.
Furthermore, introducing input asymmetry~\cite{SM} could lead to even more diverse interference phenomena.
Understanding the interplay between photon statistics and circuit symmetries
thus remains an intriguing challenge for future studies.

{\it Acknowledgements. -- }
JST FOREST Program~(JPMJFR222V);
R \& D of ICT Priority Technology~(JPMI00316);
Asahi Glass Foundation; 
JST Moonshot R \& D~(JPMJMS2066, JPMJMS226C); 
MEXT/JSPS KAKENHI~(JP25K01263).


\newpage
\mbox{ }
\newpage

\onecolumngrid
\setcounter{section}{0}
\setcounter{subsection}{0}
\setcounter{equation}{0}
\setcounter{figure}{0}
\setcounter{table}{0}

\setcounter{page}{1}

\renewcommand{\thesection}{S\arabic{section}}
\renewcommand{\thesubsection}{S\arabic{section}.\arabic{subsection}}
\renewcommand{\theequation}{S\arabic{equation}}
\renewcommand{\thefigure}{S\arabic{figure}}
\renewcommand{\thetable}{S\arabic{table}}

\section*{Supplemental Material for\\
``Statistical-noise-enhanced multi-photon interference''}

\section{Coincidence probabilities for $N = 2$}
To derive the coincidence probabilities including the mode mismatch among the input light fields,  
we explicitly treat the distinguishable inputs as occupying orthogonal modes 
which experimentally corresponds to different temporal modes, polarization, and so on. 
As shown in Fig.~\ref{fig:circuit}~(c), 
since the coincidence probabilities are insensitive to these degrees of freedom, 
the photon number operator for the $i$-th output mode is given by the sum of the intensities as 
\begin{align}
  \hat{n}^{\U{out}}_i = \sum_{j}\hat{n}^{\U{out}}_{i,j}, 
\end{align}
where the subscript $j$ represents the internal degrees of freedom,
and $\hat{n}^{\U{out}}_{i,j}=\hat{b}_{i,j}^\dagger \hat{b}_{i,j}$. 
The annihilation and creation operators satisfy $[\hat{b}_{i,j},\hat{b}_{i',j'}^\dagger]=\delta_{ii'}\delta_{jj'}$. 
Using this decomposition, the cross-terms vanish for distinguishable inputs.
We denote the number operators of the corresponding input modes by 
$\hat{n}_{i,j}=\hat{a}_{i,j}^\dagger \hat{a}_{i,j}$, with satisfying
$[\hat{a}_{i,j},\hat{a}_{i',j'}^\dagger]=\delta_{ii'}\delta_{jj'}$. 
The input and output operators are related via the matrix elements $u_{ij}$ as 
\begin{align}
\hat{b}_{i,j}^\dagger = \sum_{k}u_{ik}\hat{a}_{k,j}^\dagger, \quad 
\hat{b}_{i,j} = \sum_{k}u_{ik}^*\hat{a}_{k,j}. 
\end{align}

For $N=2$, the calculation is based on Ref.~\cite{Tsujimoto2023}. 
The coincidence probability $\PQT$ in Eq.~(\ref{eq:PQ2}) in the main text is obtained 
using $\expect{\hat{n}_{1,1}}=\expect{\hat{n}_{2,1}}=n$
and $\expect{\hat{n}_{1,j}}=\expect{\hat{n}_{2,j}}=0$ for all $j\neq 1$. 
On the other hand, $\PCT$ in Eq.~(\ref{eq:PC2}) is obtained
using $\expect{\hat{n}_{1,1}}=\expect{\hat{n}_{2,2}}=n$, 
$\expect{\hat{n}_{1,j}}=0$ for $j\neq 1$ and $\expect{\hat{n}_{2,j}}=0$ for $j\neq 2$.
To calculate the coincidence probability considering partial mode mismatch, 
we assume that the first input satisfies $\expect{\hat{n}_{1,1}}=n$,
while the second input is decomposed into $\expect{\hat{n}_{2,1}}=Mn$ and $\expect{\hat{n}_{2,2}}=(1-M)n$. 
Due to the single-mode assumption of the input light, 
the intensity correlation function is invariant under mode decomposition, 
resulting in
$\gn{2}=\expect{:\hat{n}_{2,1}^2:}/\expect{\hat{n}_{2,1}}^2
=\expect{:\hat{n}_{2,2}^2:}/\expect{\hat{n}_{2,2}}^2
=\expect{\hat{n}_{2,1}\hat{n}_{2,2}}/(\expect{\hat{n}_{2,1}}\expect{\hat{n}_{2,2}})$. 

\section{Coincidence probabilities $\PQ$ and $\PC$ for $N=3$}
The case $N=3$ can be directly calculated in the same way.
To clarify the structure of the coincidence probability $\PQ$ described by the transition matrix $U$,
we explicitly express it in terms of the matrix elements $u_{ij}$, 
considering situations where the mean intensity and intensity correlation functions are not necessarily equal.
Assuming that all input photons are created by $\hat{a}_{i,1}^\dagger$ and 
denoting $n_i=\expect{\hat{n}_{i,1}}$ and $\gn{m}_i=\expect{:\hat{n}_{i,1}^m:}/n_i^m$,
we obtain
\begin{align}
  \PQ &= |u_{11}u_{21}u_{31}|^2 n_1^3\gn{3}_1
        + |u_{12}u_{22}u_{32}|^2 n_2^3\gn{3}_2
        + |u_{13}u_{23}u_{33}|^2 n_3^3\gn{3}_3\nonumber \\
      & + |u_{11}u_{21}u_{32} + u_{11}u_{22}u_{31}+ u_{12}u_{21}u_{31}|^2 n_1^2n_2\gn{2}_1 
        + |u_{11}u_{21}u_{33} + u_{11}u_{23}u_{31}+ u_{13}u_{21}u_{31}|^2 n_1^2n_3\gn{2}_1 \nonumber \\
      & + |u_{12}u_{22}u_{31} + u_{12}u_{21}u_{32}+ u_{11}u_{22}u_{32}|^2 n_1n_2^2\gn{2}_2 
        + |u_{12}u_{22}u_{33} + u_{12}u_{23}u_{32}+ u_{13}u_{22}u_{32}|^2 n_2^2n_3\gn{2}_2 \nonumber \\
      & + |u_{13}u_{23}u_{31} + u_{13}u_{21}u_{33}+ u_{11}u_{23}u_{33}|^2 n_1n_3^2\gn{2}_3 
        + |u_{13}u_{23}u_{32} + u_{13}u_{22}u_{33}+ u_{12}u_{23}u_{33}|^2 n_2n_3^2\gn{2}_3 \nonumber \\
       & + |\U{Per}(U)|^2n_1n_2n_3,
         \label{eq:PQfull}
\end{align}
where $\U{Per}(U)$ is the permanent of matrix $U$. 
In contrast, when the photons from different input ports are completely distinguishable, 
such that the mean photon numbers are given by
$n_1=\expect{\hat{n}_{1,1}}$, $n_2=\expect{\hat{n}_{2,2}}$, and $n_3=\expect{\hat{n}_{3,3}}$ 
with no photons populating the remaining input modes, 
all interferometric effects vanish, leading to 
\begin{align}
  \PC &= v_{11}v_{21}v_{31} n_1^3\gn{3}_1
        + v_{12}v_{22}v_{32} n_2^3\gn{3}_2
        + v_{13}v_{23}v_{33} n_3^3\gn{3}_3\nonumber \\
      & + (v_{11}v_{21}v_{32} + v_{11}v_{22}v_{31}+ v_{12}v_{21}v_{31}) n_1^2n_2\gn{2}_1 
        + (v_{11}v_{21}v_{33} + v_{11}v_{23}v_{31}+ v_{13}v_{21}v_{31}) n_1^2n_3\gn{2}_1 \nonumber \\
      & + (v_{12}v_{22}v_{31} + v_{12}v_{21}v_{32}+ v_{11}v_{22}v_{32}) n_1n_2^2\gn{2}_2 
        + (v_{12}v_{22}v_{33} + v_{12}v_{23}v_{32}+ v_{13}v_{22}v_{32}) n_2^2n_3\gn{2}_2 \nonumber \\
      & + (v_{13}v_{23}v_{31} + v_{13}v_{21}v_{33}+ v_{11}v_{23}v_{33}) n_1n_3^2\gn{2}_3 
        + (v_{13}v_{23}v_{32} + v_{13}v_{22}v_{33}+ v_{12}v_{23}v_{33}) n_2n_3^2\gn{2}_3 \nonumber \\
      & + \U{Per}(V)n_1n_2n_3,  
         \label{eq:PCfull}
\end{align}
where $V$ is a matrix with elements $v_{ij}=|u_{ij}|^2$. 
When we assume symmetric inputs with identical mean photon number $n_1=n_2=n_3$, 
using Eqs.~(\ref{eq:PQfull}) and (\ref{eq:PCfull}) with matrix elements of $U\u{DFT}$ in Eqs.~(\ref{eq:DFT}), 
the coincidence probabilities in Eqs.~(\ref{eq:PQDFT}) and (\ref{eq:PCDFT}) in the main text are derived.

\section{Detailed explanation of Fig.~\ref{fig:DFT_vis}}
We explain the plot points and curves shown in Fig.~\ref{fig:DFT_vis}. 
The correlation functions for $n$-photon Fock state $\ket{n}$ are 
\begin{align}
  \gn{2}= 1-\frac{1}{n},\qquad 
  \gn{3}= \left( 1-\frac{1}{n}\right) \left( 1-\frac{2}{n} \right) .
  \label{eq:Fock}
\end{align}
As a result, from Eq.~(\ref{eq:V3}), the visibility for the $n$-photon Fock-state inputs is given by 
\begin{align}
  \Vn{3}_{\U{Fock},n} = \frac{6(1-n^{-1}) - 1}{(1-n^{-1})(1-2n^{-1}) + 6(1-n^{-1}) + 2}. 
\label{eq:VFock}
\end{align}
Using the equation,
we obtain the visibilities
$\Vn{3}_{\U{Fock},1} = -0.5$, $\Vn{3}_{\U{Fock},2} = 0.4$, and $\Vn{3}_{\U{Fock},4} \sim 0.5$ 
for the 1-, 2-, and 4-photon inputs, respectively, as depicted in Fig.~\ref{fig:DFT_vis}. 

For the Gaussian-state inputs, the upper bound on the visibility is 
\begin{align}
  \Vn{3}\u{Gauss} = \frac{6\gn{2} - 1}{(2-3\sqrt{\gn{2}})^2 +6\gn{2}+2 }, 
\end{align}
which is the green solid curve in Fig.~\ref{fig:DFT_vis}.
In the limit $\gn{2}\rightarrow \infty$, the visibility approaches a constant value, 
$\Vn{3}\u{Gauss} \rightarrow 6\gn{2}/(15\gn{2}) = 0.4$. 

If arbitrary quantum states are allowed, the visibility can span the full range from $-0.5$ to $1$. 
To see this, we consider a mixture of the vacuum, single-photon state, and two-photon state 
with probabilities $p$, $(1-p)q$ and $(1-p)(1-q)$, respectively.
$\gn{2}$ and $\gn{3}$ of the state are 
$\gn{2} = 2(1-q)(1-p)^{-1}(2-q)^{-2}$ and $0$, respectively. 
From Eq.~(\ref{eq:V3}), the visibility is
\begin{align}
  \Vn{3}\u{mix} = \frac{12(1-q)(1-p)^{-1}(2-q)^{-2} - 1}{12(1-q)(1-p)^{-1}(2-q)^{-2} + 2}.  
\end{align}
When $q=1-p$ is satisfied, 
$\Vn{3}\u{mix}$ takes $-0.5$ for $p=0$,
and approaches $1$ in the limit $p\rightarrow 1$. 
Since it is a continuous function of $p$,
$\Vn{3}\u{mix}$ can take any arbitrary value between these limits. 

\section{Coincidence probabilities with partial mode mismatch in the DFT for $N=3$}
We consider partial mode mismatch like in Fig.~\ref{fig:modemismatch},
assuming symmetric inputs with identical mean photon number $n_1=n_2=n_3=n$, 
and photon statistics $\gn{3}$ and $\gn{2}$ for all input modes. 
We first derive the coincidence probability $P\u{coinc}^{(0,M,0)}$ in Eq.~(\ref{eq:P1}) 
for $0\leq \xi \leq 1$.
In this case, $\expect{\hat{n}_{1,1}}=\expect{\hat{n}_{2,2}}=n$,
$\expect{\hat{n}_{3,2}}=Mn$ and $\expect{\hat{n}_{3,3}}=(1-M)n$ are assumed.
From the symmetry of the $3\times 3$ DFT and the photons in the different inputs,
\begin{align}
  P\u{coinc}^{(0,M,0)}
  &=  \expect{: \sum_{j,j',j''=1}^3\hat{n}^{\U{out}}_{1,j}\hat{n}^{\U{out}}_{2,j'}\hat{n}^{\U{out}}_{3,j''} :}\\
  &=  \expect{:\hat{n}^{\U{out}}_{1,1}\hat{n}^{\U{out}}_{2,1}\hat{n}^{\U{out}}_{3,1}:}
    + \expect{:\hat{n}^{\U{out}}_{1,2}\hat{n}^{\U{out}}_{2,2}\hat{n}^{\U{out}}_{3,2}:}
    + \expect{:\hat{n}^{\U{out}}_{1,3}\hat{n}^{\U{out}}_{2,3}\hat{n}^{\U{out}}_{3,3}:}\\
  & + 3 \expect{:\hat{n}^{\U{out}}_{1,2}\hat{n}^{\U{out}}_{2,1}\hat{n}^{\U{out}}_{3,1}:}
    + 3 \expect{:\hat{n}^{\U{out}}_{1,3}\hat{n}^{\U{out}}_{2,1}\hat{n}^{\U{out}}_{3,1}:}
    + 3 \expect{:\hat{n}^{\U{out}}_{1,1}\hat{n}^{\U{out}}_{2,2}\hat{n}^{\U{out}}_{3,2}:}\\
  & + 3 \expect{:\hat{n}^{\U{out}}_{1,1}\hat{n}^{\U{out}}_{2,3}\hat{n}^{\U{out}}_{3,3}:}
    + 3 \expect{:\hat{n}^{\U{out}}_{1,2}\hat{n}^{\U{out}}_{2,2}\hat{n}^{\U{out}}_{3,3}:}
    + 3 \expect{:\hat{n}^{\U{out}}_{1,2}\hat{n}^{\U{out}}_{2,3}\hat{n}^{\U{out}}_{3,3}:}\\
  & + 6 \expect{:\hat{n}^{\U{out}}_{1,1}\hat{n}^{\U{out}}_{2,2}\hat{n}^{\U{out}}_{3,3}:}.
\end{align}
Thus, the normalized coincidence probability $\tilde{P}\u{coinc}^{(0,M,0)}= P\u{coinc}^{(0,M,0)}/n^3$ is given by 
\begin{align}
  \tilde{P}\u{coinc}^{(0,M,0)}
  & = \frac{1}{27}\gn{3} + \frac{1+M^3}{27}\gn{3} + \frac{(1-M)^3}{27}\gn{3} \\
  & + \frac{1+M}{9}\gn{2} + \frac{1-M}{9}\gn{2}
    + \left( \frac{1+M^2}{9}\gn{2}+\frac{M}{9}\right)\\
  & + \frac{(1-M)^2}{9}\gn{2} + \left(\frac{1-M^2}{9}\gn{2} + \frac{M^2(1-M)}{9}\gn{3}\right)
    + \left(\frac{(1-M)^2}{9}\gn{2} + \frac{M(1-M)^2}{9}\gn{3}\right)\\
  & + \left( \frac{2M(1-M)}{9}\gn{2} + \frac{2(1-M)}{9}\right)\\
  & = \frac{1}{9}\gn{3} + \frac{2(3-M)}{9} \gn{2} + \frac{2-M}{9}. 
\end{align}

Second, we derive the coincidence probability $P\u{coinc}^{(M,1,M)}$ in Eq.~(\ref{eq:P2}) 
for $1\leq \xi \leq 2$. The probability is equivalent to $P\u{coinc}^{(1,M,M)}$. 
In this case, $\expect{\hat{n}_{1,1}}=\expect{\hat{n}_{2,1}}=n$, 
$\expect{\hat{n}_{3,1}}=Mn$ and $\expect{\hat{n}_{3,2}}=(1-M)n$ are assumed. 
\begin{align}
P\u{coinc}^{(1,M,M)}
  &= \expect{: \sum_{j,j',j''=1}^2\hat{n}^{\U{out}}_{1,j}\hat{n}^{\U{out}}_{2,j'}\hat{n}^{\U{out}}_{3,j''} :}\\
  &= \expect{:\hat{n}^{\U{out}}_{1,1}\hat{n}^{\U{out}}_{2,1}\hat{n}^{\U{out}}_{3,1}:}
    + \expect{:\hat{n}^{\U{out}}_{1,2}\hat{n}^{\U{out}}_{2,2}\hat{n}^{\U{out}}_{3,2}:}
    + 3 \expect{:\hat{n}^{\U{out}}_{1,1}\hat{n}^{\U{out}}_{2,1}\hat{n}^{\U{out}}_{3,2}:}
    + 3 \expect{:\hat{n}^{\U{out}}_{1,1}\hat{n}^{\U{out}}_{2,2}\hat{n}^{\U{out}}_{3,2}:}. 
\end{align}
As a result, we obtain
\begin{align}
\tilde{P}\u{coinc}^{(1,M,M)}
  & = \left( \frac{2+M^3}{27}\gn{3} + \frac{M}{3}\right) + \frac{(1-M)^3}{27}\gn{3}\\
  & + \frac{1-M}{9} \left( 1+2 (1+M)\gn{2} + M^2\gn{3}  \right)
    + \left( \frac{2(1-M)^2}{9}\gn{2} + \frac{M(1-M)^2}{9}\gn{3} \right)\\
  & = \frac{1}{9}\gn{3} + \frac{4(1-M)}{9} \gn{2} + \frac{1+2M}{9}. 
\end{align}

\section{Visualization of coincidence probabilities in the symmetric circuit $U\u{sym}$}
\begin{figure}[t]
 \begin{center}
   \scalebox{1.3}{\includegraphics{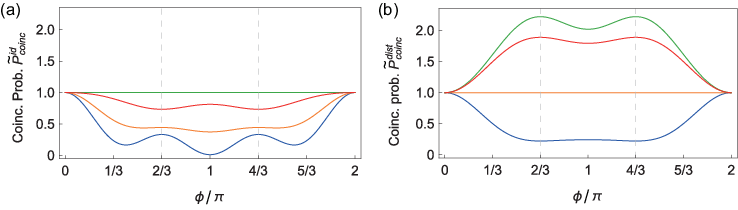}}
   \caption{
     Coincidence probabilities normalized by the input intensities as a function of phase $\phi$.
     (a) $\PQT$ for indistinguishable case. 
     (b) $\PCT$ for distinguishable case. 
     The vertical axes are fixed to the same scale
     to highlight the absolute difference in coincidence probabilities. 
     Blue: single photons. Orange: laser light. Green: thermal light. Red: optimized noise for the DFT. 
   }
    \label{fig:rawcoinc}
 \end{center}
\end{figure}
For $U\u{sym}$ in (\ref{eq:sym}), using Eqs.~(\ref{eq:PQfull}) and (\ref{eq:PCfull}) 
with assuming symmetric inputs with identical mean photon number $n_1=n_2=n_3$, 
the coincidence probabilities are given by 
\begin{align}
  \PQT (\phi) &= 3 |\alpha|^2|\beta|^4 \gn{3}
  + 6 |\beta|^2 \left| \alpha^2 + \alpha\beta + \beta^2 \right|^2 \gn{2}
                + |\alpha^3 + 3\alpha \beta^2 + 2\beta^3|^2,
                \label{eq:Pidphi} \\
  \PCT (\phi) &= 3|\alpha|^2|\beta|^4\gn{3}
  + 6 |\beta|^2\left( |\alpha|^4 + |\alpha|^2|\beta|^2 + |\beta|^4 \right) \gn{2}
  + |\alpha|^6+3|\alpha|^2|\beta|^4+2|\beta|^6. 
                \label{eq:Pdistphi}
\end{align}

We visualize the coincidence probabilities $\PQT (\phi)$ and $\PCT (\phi)$ 
in Figs.~\ref{fig:rawcoinc}~(a) and (b), respectively. 
In Fig.~\ref{fig:rawcoinc}~(a), 
the destructive interference generally suppreses the coincidence probabilities. 
For the single-photon inputs, partial constructive interference results in the 
formation of coincidence peaks at the DFT points. 
In contrast, the engineered super-Poissonian light, optimized for DFT, 
exhibits the opposite dependence on $\phi$ compared to the single photon. 
In Fig.~\ref{fig:rawcoinc}~(b), 
the coincidence probability $\PCT(\phi)$ for the single-photon inputs varies smoothly with $\phi$
due to the absence of interference. 
Consequently, an inversion of the fringe pattern with respect to $\PQT(\phi)$ occurs,
leading to a pronounced change in the visibility, as shown in Fig.~\ref{fig:sym_vis}.
For the thermal light and optimized noise,
the coincidence probabilities are enhanced compared to the laser light 
due to the classical bunching effect originating from $\gn{2}>1$ and $\gn{3}>1$. 
We note that, for the thermal light,
the bunching enhancement and the destructive interference effect exactly cancel each other,
rendering the coincidence probability $\PQT(\phi)$ independent of $\phi$ as in Fig.~\ref{fig:rawcoinc}~(a).

\section{Numerical analysis of visibilities for Fock-state and classical-light inputs}

\begin{figure}[t]
 \begin{center}
   \scalebox{1.3}{\includegraphics{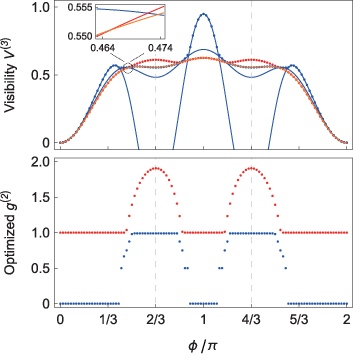}}
   \caption{
     Numerical simulation of maximized visibilities
     for optimized noise~(red points) calculated from Eq.~(\ref{eq:Vcl})
     and $n$-photon Fock-state inputs~(blue points) from Eq.~(\ref{eq:VFock}). 
     The solid blue curves correspond to single-photon~($n=1$) and three-photon~($n=3$) inputs,
     while the solid orange curve indicates laser-light inputs. 
     The lower figure shows the corresponding optimal $\gn{2}$ values for the maximization. 
     In the inset, the sollid red curve represents the visibility obtained using Eq.~(\ref{eq:Vcl})
     with the statistics~($\gn{2}\sim 1.13$) fixed at the optimal condition for $\phi/\pi=0.471$.
   }
    \label{fig:opt}
 \end{center}
\end{figure}
In Fig.~\ref{fig:sym_vis} in the main text,
the red curve for showing the classical advantage was obtained 
for the super Poissonian light optimized for DFT, with $\gn{2}\sim 1.9$ and $\gn{3} \sim 3.6$. 
More precisely, for each value of $\phi$,
a higher visibility can be achieved by using classical light with an optimized $\gn{2}$
at the region of $\gn{2}\geq 1$. 
Using Eqs.~(\ref{eq:Pidphi}), (\ref{eq:Pdistphi}) and $\gn{3}=(\gn{2})^2$, 
we numerically maximize the visibility $\Vn{3}(\phi)=1-\PQT (\phi)/\PCT (\phi)$ 
with respect to $\gn{2}$ for each value of $\phi$. 
Simultaneously, using Eq.~(\ref{eq:VFock}),
we maximize the absolute value of visibility for the Fock-state inputs. 

The result is shown in Fig.~\ref{fig:opt}.
For the Fock-state inputs, beyond the single-photon peak~($\phi \sim \pi/3$),
utilizing larger Fock states allows for maintaining higher visibility
as the phase approaches the crossover with the laser limit.
Conversely, regarding classical light, the visibility surpasses the laser limit near this crossover.
From this point, optimizing $\gn{2}$ for each $\phi$ maximizes the visibility
until it eventually converges back to the laser level. 

While the crossover points of visibilities for the Fock-state, super-Poissonian, and laser lights are nearly the same, 
we numerically investigated the region, where the visibilities for the Fock-state and super-Poissonian inputs 
surpass the visibility for the laser-light inputs, based on the following strategy.
We first identified the analytical crossover point between the Fock state and the laser as $\phi = \pi/2$
by considering the $n \rightarrow \infty$ limit, with the visibility of $\sim 0.560$. 
At this phase, the optimized noise visibility is found to $\sim 0.567$ at $\gn{2}\sim 1.39$ 
which slightly exceeds the laser benchmark.
Therefore, we explored the phase region slightly smaller than $\pi/2$,
where the visibility of the Fock-state inputs increases while that of the noise inputs decreases.
As a result, we successfully identified a narrow regime for $n \geq 3$ 
where both quantum and noise inputs marginally exceed the laser limit,
as shown in inset of Fig.~\ref{fig:opt}.

\section{Coincidence probabilities for arbitrary $N$}
The coincidence probabilities for $N$ can be derived
using a method similar to that employed in Ref.~\cite{Tichy2010}. 
For the calculation, we define the tuple $\vect{s}=(s_1,\ldots, s_N)$ of operator exponents that satisfies 
$\sum s_i = N$, with non-negative integers $s_i$. 
In addition, we define $\vect{d}(\vect{s})$ as 
\begin{align}
\vect{d}(\vect{s})=\oplus_{j=1}^N \oplus_{k=1}^{s_j}(j), 
\end{align}
which represents the mode assignment tuple of length $N$,
constructed by repeating the mode index $j$ exactly $s_j$ times. 
For example, the configuration $\vect{s}=(2,0,1)$, maps to the mode assignment $\vect{d}(\vect{s})=(1,1,3)$. 
In the calculation of $\PQ$ under perfect mode matching, 
we assume that all input photons are created by $\hat{a}_{i,1}^\dagger$. 
By omitting the internal mode label $1$ for readability, we obtain
\begin{align}
  \PQ &= \expect{
        \prod_{i}
        \left(
        \sum_{k} u_{ik}\hat{a}_{k}^\dagger
        \right)
        \,
        \prod_{i'}
        \left(
        \sum_{k'} u_{i'k'}^* \hat{a}_{k'}
        \right)
        }\\
      &=
        \expect{
        \left( \sum_{\vect{s}} \frac{\U{Per}(U_{\vect{d}(\vect{s})})}{\prod_{i} s_i!}
        \,
        \prod_{i} \hat{a}_{i}^{\dagger s_i} \right)
        \left( \sum_{\vect{s}'} \frac{\U{Per}(U^*_{\vect{d}(\vect{s}')})}{\prod_{i} s'_i!}
        \,
        \prod_{i} \hat{a}_{i}^{s'_i} \right)
        }
        \label{eq:phase} \\
      &=
          \sum_{\vect{s}}
          \left|
          \frac{\U{Per} (U_{\vect{d}(\vect{s})})}{\prod_{i} s_i!}
          \right|^2
          \expect{ \prod_{i} \hat{a}_i^{\dagger s_i}\hat{a}_i^{s_i} }
        \label{eq:statistics}\\
      &=
        \sum_{\vect{s}}
        \left|
        \frac{\U{Per} (U_{\vect{d}(\vect{s})})}{\prod_{i} s_i!}
        \right|^2
        \prod_i n_i^{s_i} \gn{s_i}_i, 
        \label{eq:final}
\end{align}
where $U_{\vect{d}(\vect{s})}$ denotes the $N \times N$ matrix
constructed by selecting the columns of $U$ corresponding to the indices in $\vect{d}(\vect{s})$,
whose matrix elements are given by $[U_{\vect{d}(\vect{s})}]_{jk} = u_{j, d_k(\vect{s})}$.
From Eq.~(\ref{eq:phase}) to (\ref{eq:statistics}), 
applying the assumption of phase independence of the inputs restricts the sum to $\vect{s}=\vect{s}'$.
From Eq.~(\ref{eq:statistics}) to (\ref{eq:final}), 
due to the statistical independence, the remaining expectation value factorizes, 
resulting in an expression including the mean photon number and intensity correlation functions. 

In contrast, the coincidence probability $\PC$ under complete mode mismatch is given by 
\begin{align}
  \PC &= 
        \sum_{\vect{s}}
        \frac{\U{Per} (V_{\vect{d}(\vect{s})})}{\prod_{i} s_i!}
        \prod_i n_i^{s_i} \gn{s_i}_i, 
\end{align}
where $V_{\vect{d}(\vect{s})}$ is the matrix defined by elements $[V_{\vect{d}(\vect{s})}]_{jk} = |u_{j, d_k(\vect{s})}|^2$.


\begin{thebibliography}{30}
\expandafter\ifx\csname natexlab\endcsname\relax\def\natexlab#1{#1}\fi
\expandafter\ifx\csname bibnamefont\endcsname\relax
  \def\bibnamefont#1{#1}\fi
\expandafter\ifx\csname bibfnamefont\endcsname\relax
  \def\bibfnamefont#1{#1}\fi
\expandafter\ifx\csname citenamefont\endcsname\relax
  \def\citenamefont#1{#1}\fi
\expandafter\ifx\csname url\endcsname\relax
  \def\url#1{\texttt{#1}}\fi
\expandafter\ifx\csname urlprefix\endcsname\relax\def\urlprefix{URL }\fi
\providecommand{\bibinfo}[2]{#2}
\providecommand{\eprint}[2][]{\url{#2}}

\bibitem[{\citenamefont{Hong et~al.}(1987)\citenamefont{Hong, Ou, and
  Mandel}}]{Hong1987}
\bibinfo{author}{\bibfnamefont{C.~K.} \bibnamefont{Hong}},
  \bibinfo{author}{\bibfnamefont{Z.~Y.} \bibnamefont{Ou}}, \bibnamefont{and}
  \bibinfo{author}{\bibfnamefont{L.}~\bibnamefont{Mandel}},
  \emph{\bibinfo{title}{Measurement of subpicosecond time intervals between two
  photons by interference}}, \bibinfo{journal}{Phys. Rev. Lett.}
  \textbf{\bibinfo{volume}{59}}, \bibinfo{pages}{2044} (\bibinfo{year}{1987}).

\bibitem[{\citenamefont{Lim and Beige}(2005)}]{Lim2005}
\bibinfo{author}{\bibfnamefont{Y.~L.} \bibnamefont{Lim}} \bibnamefont{and}
  \bibinfo{author}{\bibfnamefont{A.}~\bibnamefont{Beige}},
  \emph{\bibinfo{title}{{Generalized Hong--Ou--Mandel experiments with bosons
  and fermions}}}, \bibinfo{journal}{New Journal of Physics}
  \textbf{\bibinfo{volume}{7}}, \bibinfo{pages}{155} (\bibinfo{year}{2005}).

\bibitem[{\citenamefont{Tichy et~al.}(2010)\citenamefont{Tichy, Tiersch,
  de~Melo, Mintert, and Buchleitner}}]{Tichy2010}
\bibinfo{author}{\bibfnamefont{M.~C.} \bibnamefont{Tichy}},
  \bibinfo{author}{\bibfnamefont{M.}~\bibnamefont{Tiersch}},
  \bibinfo{author}{\bibfnamefont{F.}~\bibnamefont{de~Melo}},
  \bibinfo{author}{\bibfnamefont{F.}~\bibnamefont{Mintert}}, \bibnamefont{and}
  \bibinfo{author}{\bibfnamefont{A.}~\bibnamefont{Buchleitner}},
  \emph{\bibinfo{title}{Zero-transmission law for multiport beam splitters}},
  \bibinfo{journal}{Physical review letters} \textbf{\bibinfo{volume}{104}},
  \bibinfo{pages}{220405} (\bibinfo{year}{2010}).

\bibitem[{\citenamefont{Spagnolo et~al.}(2013)\citenamefont{Spagnolo, Vitelli,
  Aparo, Mataloni, Sciarrino, Crespi, Ramponi, and Osellame}}]{Spagnolo2013-2}
\bibinfo{author}{\bibfnamefont{N.}~\bibnamefont{Spagnolo}},
  \bibinfo{author}{\bibfnamefont{C.}~\bibnamefont{Vitelli}},
  \bibinfo{author}{\bibfnamefont{L.}~\bibnamefont{Aparo}},
  \bibinfo{author}{\bibfnamefont{P.}~\bibnamefont{Mataloni}},
  \bibinfo{author}{\bibfnamefont{F.}~\bibnamefont{Sciarrino}},
  \bibinfo{author}{\bibfnamefont{A.}~\bibnamefont{Crespi}},
  \bibinfo{author}{\bibfnamefont{R.}~\bibnamefont{Ramponi}}, \bibnamefont{and}
  \bibinfo{author}{\bibfnamefont{R.}~\bibnamefont{Osellame}},
  \emph{\bibinfo{title}{Three-photon bosonic coalescence in an integrated
  tritter}} (\bibinfo{year}{2013}).

\bibitem[{\citenamefont{Crespi}(2015)}]{Crespi2015}
\bibinfo{author}{\bibfnamefont{A.}~\bibnamefont{Crespi}},
  \emph{\bibinfo{title}{Suppression laws for multiparticle interference in
  sylvester interferometers}}, \bibinfo{journal}{Phys. Rev. A}
  \textbf{\bibinfo{volume}{91}}, \bibinfo{pages}{013811}
  (\bibinfo{year}{2015}).

\bibitem[{\citenamefont{Shchesnovich}(2016)}]{Shchesnovich2016}
\bibinfo{author}{\bibfnamefont{V.~S.} \bibnamefont{Shchesnovich}},
  \emph{\bibinfo{title}{Universality of generalized bunching and efficient
  assessment of boson sampling}}, \bibinfo{journal}{Phys. Rev. Lett.}
  \textbf{\bibinfo{volume}{116}}, \bibinfo{pages}{123601}
  (\bibinfo{year}{2016}).

\bibitem[{\citenamefont{Crespi et~al.}(2016)\citenamefont{Crespi, Osellame,
  Ramponi, Bentivegna, Flamini, Spagnolo, Viggianiello, Innocenti, Mataloni,
  and Sciarrino}}]{Crespi2016}
\bibinfo{author}{\bibfnamefont{A.}~\bibnamefont{Crespi}},
  \bibinfo{author}{\bibfnamefont{R.}~\bibnamefont{Osellame}},
  \bibinfo{author}{\bibfnamefont{R.}~\bibnamefont{Ramponi}},
  \bibinfo{author}{\bibfnamefont{M.}~\bibnamefont{Bentivegna}},
  \bibinfo{author}{\bibfnamefont{F.}~\bibnamefont{Flamini}},
  \bibinfo{author}{\bibfnamefont{N.}~\bibnamefont{Spagnolo}},
  \bibinfo{author}{\bibfnamefont{N.}~\bibnamefont{Viggianiello}},
  \bibinfo{author}{\bibfnamefont{L.}~\bibnamefont{Innocenti}},
  \bibinfo{author}{\bibfnamefont{P.}~\bibnamefont{Mataloni}}, \bibnamefont{and}
  \bibinfo{author}{\bibfnamefont{F.}~\bibnamefont{Sciarrino}},
  \emph{\bibinfo{title}{{Suppression law of quantum states in a 3D photonic
  fast Fourier transform chip}}}, \bibinfo{journal}{Nature communications}
  \textbf{\bibinfo{volume}{7}}, \bibinfo{pages}{10469} (\bibinfo{year}{2016}).

\bibitem[{\citenamefont{Dittel et~al.}(2018)\citenamefont{Dittel, Dufour,
  Walschaers, Weihs, Buchleitner, and Keil}}]{Dittel2018}
\bibinfo{author}{\bibfnamefont{C.}~\bibnamefont{Dittel}},
  \bibinfo{author}{\bibfnamefont{G.}~\bibnamefont{Dufour}},
  \bibinfo{author}{\bibfnamefont{M.}~\bibnamefont{Walschaers}},
  \bibinfo{author}{\bibfnamefont{G.}~\bibnamefont{Weihs}},
  \bibinfo{author}{\bibfnamefont{A.}~\bibnamefont{Buchleitner}},
  \bibnamefont{and} \bibinfo{author}{\bibfnamefont{R.}~\bibnamefont{Keil}},
  \emph{\bibinfo{title}{Totally destructive many-particle interference}},
  \bibinfo{journal}{Phys. Rev. Lett.} \textbf{\bibinfo{volume}{120}},
  \bibinfo{pages}{240404} (\bibinfo{year}{2018}).

\bibitem[{\citenamefont{Yamazaki et~al.}(2023)\citenamefont{Yamazaki, Ikuta,
  and Yamamoto}}]{Yamazaki2023arxiv}
\bibinfo{author}{\bibfnamefont{T.}~\bibnamefont{Yamazaki}},
  \bibinfo{author}{\bibfnamefont{R.}~\bibnamefont{Ikuta}}, \bibnamefont{and}
  \bibinfo{author}{\bibfnamefont{T.}~\bibnamefont{Yamamoto}},
  \emph{\bibinfo{title}{{Stabilizer formalism in linear optics and application
  to Bell-state discrimination}}}, \bibinfo{journal}{arXiv preprint
  arXiv:2301.06551}  (\bibinfo{year}{2023}).

\bibitem[{\citenamefont{Shchesnovich}(2015)}]{Shchesnovich2015}
\bibinfo{author}{\bibfnamefont{V.~S.} \bibnamefont{Shchesnovich}},
  \emph{\bibinfo{title}{Partial indistinguishability theory for multiphoton
  experiments in multiport devices}}, \bibinfo{journal}{Phys. Rev. A}
  \textbf{\bibinfo{volume}{91}}, \bibinfo{pages}{013844}
  (\bibinfo{year}{2015}).

\bibitem[{\citenamefont{Tichy}(2015)}]{Tichy2015}
\bibinfo{author}{\bibfnamefont{M.~C.} \bibnamefont{Tichy}},
  \emph{\bibinfo{title}{Sampling of partially distinguishable bosons and the
  relation to the multidimensional permanent}}, \bibinfo{journal}{Physical
  Review A} \textbf{\bibinfo{volume}{91}}, \bibinfo{pages}{022316}
  (\bibinfo{year}{2015}).

\bibitem[{\citenamefont{Karczewski et~al.}(2019)\citenamefont{Karczewski,
  Pisarczyk, and Kurzy\ifmmode~\acute{n}\else \'{n}\fi{}ski}}]{Karczewski2019}
\bibinfo{author}{\bibfnamefont{M.}~\bibnamefont{Karczewski}},
  \bibinfo{author}{\bibfnamefont{R.}~\bibnamefont{Pisarczyk}},
  \bibnamefont{and}
  \bibinfo{author}{\bibfnamefont{P.}~\bibnamefont{Kurzy\ifmmode~\acute{n}\else
  \'{n}\fi{}ski}}, \emph{\bibinfo{title}{{Genuine multipartite
  indistinguishability and its detection via the generalized Hong-Ou-Mandel
  effect}}}, \bibinfo{journal}{Phys. Rev. A} \textbf{\bibinfo{volume}{99}},
  \bibinfo{pages}{042102} (\bibinfo{year}{2019}).

\bibitem[{\citenamefont{Shchesnovich}(2022)}]{Shchesnovich2022}
\bibinfo{author}{\bibfnamefont{V.}~\bibnamefont{Shchesnovich}},
  \emph{\bibinfo{title}{Distinguishability in quantum interference with
  multimode squeezed states}}, \bibinfo{journal}{Physical Review A}
  \textbf{\bibinfo{volume}{105}}, \bibinfo{pages}{063703}
  (\bibinfo{year}{2022}).

\bibitem[{\citenamefont{Annoni and Wein}(2025)}]{Annoni2025}
\bibinfo{author}{\bibfnamefont{E.}~\bibnamefont{Annoni}} \bibnamefont{and}
  \bibinfo{author}{\bibfnamefont{S.~C.} \bibnamefont{Wein}},
  \emph{\bibinfo{title}{Incoherent behavior of partially distinguishable
  photons}}, \bibinfo{journal}{arXiv preprint arXiv:2502.05047}
  (\bibinfo{year}{2025}).

\bibitem[{\citenamefont{Seron et~al.}(2023)\citenamefont{Seron, Novo, and
  Cerf}}]{Seron2023}
\bibinfo{author}{\bibfnamefont{B.}~\bibnamefont{Seron}},
  \bibinfo{author}{\bibfnamefont{L.}~\bibnamefont{Novo}}, \bibnamefont{and}
  \bibinfo{author}{\bibfnamefont{N.~J.} \bibnamefont{Cerf}},
  \emph{\bibinfo{title}{Boson bunching is not maximized by indistinguishable
  particles}}, \bibinfo{journal}{Nature Photonics}
  \textbf{\bibinfo{volume}{17}}, \bibinfo{pages}{702} (\bibinfo{year}{2023}).

\bibitem[{\citenamefont{Pioge et~al.}(2023)\citenamefont{Pioge, Seron, Novo,
  and Cerf}}]{Pioge2023}
\bibinfo{author}{\bibfnamefont{L.}~\bibnamefont{Pioge}},
  \bibinfo{author}{\bibfnamefont{B.}~\bibnamefont{Seron}},
  \bibinfo{author}{\bibfnamefont{L.}~\bibnamefont{Novo}}, \bibnamefont{and}
  \bibinfo{author}{\bibfnamefont{N.~J.} \bibnamefont{Cerf}},
  \emph{\bibinfo{title}{Anomalous bunching of nearly indistinguishable
  bosons}}, \bibinfo{journal}{arXiv preprint arXiv:2308.12226}
  (\bibinfo{year}{2023}).

\bibitem[{\citenamefont{Glauber}(1963)}]{Glauber1963}
\bibinfo{author}{\bibfnamefont{R.~J.} \bibnamefont{Glauber}},
  \emph{\bibinfo{title}{The quantum theory of optical coherence}},
  \bibinfo{journal}{Phys. Rev.} \textbf{\bibinfo{volume}{130}},
  \bibinfo{pages}{2529} (\bibinfo{year}{1963}).

\bibitem[{\citenamefont{Mandel and Wolf}(1995)}]{Mandel1995}
\bibinfo{author}{\bibfnamefont{L.}~\bibnamefont{Mandel}} \bibnamefont{and}
  \bibinfo{author}{\bibfnamefont{E.}~\bibnamefont{Wolf}},
  \emph{\bibinfo{title}{Optical coherence and quantum optics}}
  (\bibinfo{publisher}{Cambridge university press}, \bibinfo{year}{1995}).

\bibitem[{\citenamefont{Klyshko}(1996)}]{Klyshko1996}
\bibinfo{author}{\bibfnamefont{D.}~\bibnamefont{Klyshko}},
  \emph{\bibinfo{title}{The nonclassical light}},
  \bibinfo{journal}{Physics-Uspekhi} \textbf{\bibinfo{volume}{39}},
  \bibinfo{pages}{573} (\bibinfo{year}{1996}).

\bibitem[{\citenamefont{Santori et~al.}(2002)\citenamefont{Santori, Fattal,
  Vu{\v{c}}kovi{\'c}, Solomon, and Yamamoto}}]{Santori2002}
\bibinfo{author}{\bibfnamefont{C.}~\bibnamefont{Santori}},
  \bibinfo{author}{\bibfnamefont{D.}~\bibnamefont{Fattal}},
  \bibinfo{author}{\bibfnamefont{J.}~\bibnamefont{Vu{\v{c}}kovi{\'c}}},
  \bibinfo{author}{\bibfnamefont{G.~S.} \bibnamefont{Solomon}},
  \bibnamefont{and} \bibinfo{author}{\bibfnamefont{Y.}~\bibnamefont{Yamamoto}},
  \emph{\bibinfo{title}{Indistinguishable photons from a single-photon
  device}}, \bibinfo{journal}{nature} \textbf{\bibinfo{volume}{419}},
  \bibinfo{pages}{594} (\bibinfo{year}{2002}).

\bibitem[{\citenamefont{Tsujimoto et~al.}(2017)\citenamefont{Tsujimoto,
  Sugiura, Tanaka, Ikuta, Miki, Yamashita, Terai, Fujiwara, Yamamoto, Koashi
  et~al.}}]{Tsujimoto2017}
\bibinfo{author}{\bibfnamefont{Y.}~\bibnamefont{Tsujimoto}},
  \bibinfo{author}{\bibfnamefont{Y.}~\bibnamefont{Sugiura}},
  \bibinfo{author}{\bibfnamefont{M.}~\bibnamefont{Tanaka}},
  \bibinfo{author}{\bibfnamefont{R.}~\bibnamefont{Ikuta}},
  \bibinfo{author}{\bibfnamefont{S.}~\bibnamefont{Miki}},
  \bibinfo{author}{\bibfnamefont{T.}~\bibnamefont{Yamashita}},
  \bibinfo{author}{\bibfnamefont{H.}~\bibnamefont{Terai}},
  \bibinfo{author}{\bibfnamefont{M.}~\bibnamefont{Fujiwara}},
  \bibinfo{author}{\bibfnamefont{T.}~\bibnamefont{Yamamoto}},
  \bibinfo{author}{\bibfnamefont{M.}~\bibnamefont{Koashi}},
  \bibnamefont{et~al.}, \emph{\bibinfo{title}{High visibility hong-ou-mandel
  interference via a time-resolved coincidence measurement}},
  \bibinfo{journal}{Optics Express} \textbf{\bibinfo{volume}{25}},
  \bibinfo{pages}{12069} (\bibinfo{year}{2017}).

\bibitem[{\citenamefont{Ollivier et~al.}(2021)\citenamefont{Ollivier, Thomas,
  Wein, de~Buy~Wenniger, Coste, Loredo, Somaschi, Harouri, Lemaitre, Sagnes
  et~al.}}]{Ollivier2021}
\bibinfo{author}{\bibfnamefont{H.}~\bibnamefont{Ollivier}},
  \bibinfo{author}{\bibfnamefont{S.~E.} \bibnamefont{Thomas}},
  \bibinfo{author}{\bibfnamefont{S.~C.} \bibnamefont{Wein}},
  \bibinfo{author}{\bibfnamefont{I.~M.} \bibnamefont{de~Buy~Wenniger}},
  \bibinfo{author}{\bibfnamefont{N.}~\bibnamefont{Coste}},
  \bibinfo{author}{\bibfnamefont{J.~C.} \bibnamefont{Loredo}},
  \bibinfo{author}{\bibfnamefont{N.}~\bibnamefont{Somaschi}},
  \bibinfo{author}{\bibfnamefont{A.}~\bibnamefont{Harouri}},
  \bibinfo{author}{\bibfnamefont{A.}~\bibnamefont{Lemaitre}},
  \bibinfo{author}{\bibfnamefont{I.}~\bibnamefont{Sagnes}},
  \bibnamefont{et~al.}, \emph{\bibinfo{title}{Hong-ou-mandel interference with
  imperfect single photon sources}}, \bibinfo{journal}{Phys. Rev. Lett.}
  \textbf{\bibinfo{volume}{126}}, \bibinfo{pages}{063602}
  (\bibinfo{year}{2021}).

\bibitem[{\citenamefont{Tsujimoto et~al.}(2023)\citenamefont{Tsujimoto, Ikuta,
  Wakui, Kobayashi, and Fujiwara}}]{Tsujimoto2023}
\bibinfo{author}{\bibfnamefont{Y.}~\bibnamefont{Tsujimoto}},
  \bibinfo{author}{\bibfnamefont{R.}~\bibnamefont{Ikuta}},
  \bibinfo{author}{\bibfnamefont{K.}~\bibnamefont{Wakui}},
  \bibinfo{author}{\bibfnamefont{T.}~\bibnamefont{Kobayashi}},
  \bibnamefont{and} \bibinfo{author}{\bibfnamefont{M.}~\bibnamefont{Fujiwara}},
  \emph{\bibinfo{title}{Quantum state tomography of qudits via {Hong-Ou-Mandel}
  interference}}, \bibinfo{journal}{Phys. Rev. Applied}
  \textbf{\bibinfo{volume}{19}}, \bibinfo{pages}{014008}
  (\bibinfo{year}{2023}).

\bibitem[{\citenamefont{Ou et~al.}(1987)\citenamefont{Ou, Hong, and
  Mandel}}]{Ou1987}
\bibinfo{author}{\bibfnamefont{Z.}~\bibnamefont{Ou}},
  \bibinfo{author}{\bibfnamefont{C.}~\bibnamefont{Hong}}, \bibnamefont{and}
  \bibinfo{author}{\bibfnamefont{L.}~\bibnamefont{Mandel}},
  \emph{\bibinfo{title}{Relation between input and output states for a beam
  splitter}}, \bibinfo{journal}{Optics communications}
  \textbf{\bibinfo{volume}{63}}, \bibinfo{pages}{118} (\bibinfo{year}{1987}).

\bibitem[{\citenamefont{Ikuta}(2022)}]{Ikuta2022}
\bibinfo{author}{\bibfnamefont{R.}~\bibnamefont{Ikuta}},
  \emph{\bibinfo{title}{Wave-particle duality of light appearing in an
  intensity interferometric scenario}}, \bibinfo{journal}{Optics Express}
  \textbf{\bibinfo{volume}{30}}, \bibinfo{pages}{46972} (\bibinfo{year}{2022}).

\bibitem[{\citenamefont{Rarity et~al.}(2005)\citenamefont{Rarity, Tapster, and
  Loudon}}]{Rarity2005}
\bibinfo{author}{\bibfnamefont{J.}~\bibnamefont{Rarity}},
  \bibinfo{author}{\bibfnamefont{P.}~\bibnamefont{Tapster}}, \bibnamefont{and}
  \bibinfo{author}{\bibfnamefont{R.}~\bibnamefont{Loudon}},
  \emph{\bibinfo{title}{Non-classical interference between independent
  sources}}, \bibinfo{journal}{Journal of Optics B: Quantum and Semiclassical
  Optics} \textbf{\bibinfo{volume}{7}}, \bibinfo{pages}{S171}
  (\bibinfo{year}{2005}).

\bibitem[{\citenamefont{{See Supplemental Material.}}()}]{SM}
\bibinfo{author}{\bibnamefont{{See Supplemental Material.}}}

\bibitem[{\citenamefont{Loudon}(2000)}]{Loudon2000}
\bibinfo{author}{\bibfnamefont{R.}~\bibnamefont{Loudon}},
  \emph{\bibinfo{title}{The quantum theory of light}} (\bibinfo{publisher}{OUP
  Oxford}, \bibinfo{year}{2000}).

\bibitem[{\citenamefont{Menssen et~al.}(2017)\citenamefont{Menssen, Jones,
  Metcalf, Tichy, Barz, Kolthammer, and Walmsley}}]{Menssen2017}
\bibinfo{author}{\bibfnamefont{A.~J.} \bibnamefont{Menssen}},
  \bibinfo{author}{\bibfnamefont{A.~E.} \bibnamefont{Jones}},
  \bibinfo{author}{\bibfnamefont{B.~J.} \bibnamefont{Metcalf}},
  \bibinfo{author}{\bibfnamefont{M.~C.} \bibnamefont{Tichy}},
  \bibinfo{author}{\bibfnamefont{S.}~\bibnamefont{Barz}},
  \bibinfo{author}{\bibfnamefont{W.~S.} \bibnamefont{Kolthammer}},
  \bibnamefont{and} \bibinfo{author}{\bibfnamefont{I.~A.}
  \bibnamefont{Walmsley}}, \emph{\bibinfo{title}{Distinguishability and
  many-particle interference}}, \bibinfo{journal}{Physical review letters}
  \textbf{\bibinfo{volume}{118}}, \bibinfo{pages}{153603}
  (\bibinfo{year}{2017}).

\bibitem[{\citenamefont{Hotter et~al.}(2025)\citenamefont{Hotter, Henke, van
  Diepen, Lodahl, and S{\o}rensen}}]{Hotter2025}
\bibinfo{author}{\bibfnamefont{C.}~\bibnamefont{Hotter}},
  \bibinfo{author}{\bibfnamefont{C.}~\bibnamefont{Henke}},
  \bibinfo{author}{\bibfnamefont{C.~J.} \bibnamefont{van Diepen}},
  \bibinfo{author}{\bibfnamefont{P.}~\bibnamefont{Lodahl}}, \bibnamefont{and}
  \bibinfo{author}{\bibfnamefont{A.~S.} \bibnamefont{S{\o}rensen}},
  \emph{\bibinfo{title}{A quantum non-gaussianity criterion based on photon
  correlations $g^{(2)}$ and $g^{(3)}$}}, \bibinfo{journal}{arXiv preprint
  arXiv:2511.08488}  (\bibinfo{year}{2025}).

\end{thebibliography}
\end{document}